\documentstyle[12pt]{article}
\newcommand{\be}{\begin{equation}}
\newcommand{\ee}{\end{equation}}
\newcommand{\rref}[1]{(\ref{#1})}
\begin{document}

\begin{flushright}
RI-98-8 \\
hep-th/9810180 \\
October 1998\\
\end{flushright}

\vspace{.8cm}

\begin{center}
{\huge On the Exotic Phases of M-theory}

\vspace{1.5cm}

{\large Riccardo Argurio}

\vspace{.4cm}

{\it Racah Institute of Physics}\\
{\it The Hebrew University}\\
{\it Jerusalem, 91904, Israel}\\
{Email: \tt argurio@cc.huji.ac.il}\\

\vspace{.4cm}

{\large and}

\vspace{.4cm}

{\large  Laurent Houart}\footnote{Address as of December 1st, 1998: Theoretical 
Physics Group, Imperial College, London SW7 2BZ, U.K.}
\addtocounter{footnote}{-1}\\
\vspace{.4cm}

{\it Centro de Estudios Cient\'{\i}ficos de Santiago}\\
{\it Casilla 16443, Santiago, Chile}\\
{Email: \tt laurent@einstein.cecs.cl}\\

\end{center}

\vspace{1.5cm}

\begin{abstract}

We study aspects of the new phases of M-theory recently
conjectured using generalised dualities such as timelike T-duality. 
Our focus is on brane solutions. We derive the intersection
rules in a general framework and then specialise to the new phases
of M-theory. We discuss under which conditions a configuration with
several branes leads to a regular extremal black hole under compactification.
We point out that the entropy seems not to be constant when the radius
of the physical timelike direction is varied. This could be interpreted
as a non-conservation of the entropy (and the mass) under at least
some of the new dualities.

\end{abstract}

\newpage
\section{Introduction}
The study of dualities \cite{huto,witt,hora,tasi,deuxb} relating the consistent superstring 
theories led to the picture that these theories are descriptions of different phases, 
different regimes of a unique theory called M-theory. 
M-theory can take very different forms according to 
the limit one is considering. 
These phases differ by the presence or not of a gauge group,
by the amount of supersymmetry and even by the dimensionality of the 
target space-time.
The typical example is the opening up  of a new dimension in the strong coupling  
limit of ten dimensional IIA theory which corresponds to
11 dimensional supergravity \cite{witt}.

All these theories are formulated in a space-time with Lorentzian 
signature with one time 
coordinate, and the ``orthodox" dualities do not affect the time 
direction which is supposed to be non-compact. 
Recently, timelike compactifications of M-theory and type II superstring
theories on Lorentzian tori $T^{n,1}$ with $n$ spatial circles and one timelike 
circle have been  considered  \cite{huju,crem,hul1}. 
The limits in which various cycles degenerate were studied.
It was shown \cite{hul1}  that the type IIA (resp. IIB)  theory 
on a timelike circle gives, in the limit in which the circle shrinks to zero size,  
a T-dual theory in 9+1 dimensions   called ${\rm IIB}^*$   (resp. ${\rm IIA}^*$) 
characterised by the fact that the kinetic terms  of the RR fields have
the opposite (i.e. wrong) sign\footnote{Aspects of timelike T-duality have 
been previously  discussed in a slightly different context in \cite{moor}.}. 
The ``exotic"  dualities in which timelike directions are involved 
have been further analysed in \cite{hul2}.
For example, the S-dual theory of ${\rm IIB}^*$  is a 9+1 theory called 
${\rm IIB}^{\prime}$ characterised by a NSNS three form with a kinetic term 
of the wrong sign, implying the existence  of an Euclidean fundamental string. 
Using the Buscher approach \cite{busc}, it was furthermore shown in  \cite{hul2} 
that  performing a T-duality  in a string theory characterised by an Euclidean world-sheet maps  
the theory compactified  on a spacelike (resp. timelike) circle $R$ onto a theory
compactified on a timelike (resp. spacelike) circle of radius $1/R$ changing thus 
the space time signature of the theory. Also the strong coupling of type ${\rm IIA}^*$ 
appears to be a 9+2 supergravity called  ${\rm M}^*$ and characterised by the kinetic 
term of the  four form field strength having the wrong sign \cite{hul2}.
Starting with, say, 
the usual IIA theory and performing  a chain of  T-dualities in various space 
or timelike directions and S dualities, one generates new ``exotic" phases of M-theory 
corresponding  to type IIA-like supergravities in  10+0, 9+1,  8+2, 6+4, 5+5,  4+6, 2+8, 
0+10  dimensions, type IIB-like supergravities in 9+1, 7+3,  5+5,  3+7, 
1+9 dimensions and eleven dimensional supergravities with signature 10+1, 
9+2, 6+5, 5+6, 2+9 and 1+10.
The action of all these theories differs from the usual supergravity actions by 
the fact that the signs of some of the kinetic terms are reversed \cite{hul2}. 
All of these `new' theories present however pathologies which are absent
in the usual phases of M-theory, most notably ghost fields,
and a better understanding of these phases and their relevance is certainly worthwhile. 
In a recent work \cite{huku}, the study of solitons of these theories was initiated
and a classification of all the $p$-brane 
type solutions preserving $1/2$ of the supersymmetry 
was given. Note that some of these solitons had already been considered
in the past \cite{duff} from the point of view of their supersymmetric 
world-volume action.

In this letter we propose to investigate further these new phases of M-theory,
focusing on their brane solutions. We first derive the  
intersection rules \cite{tsey,kast,vand} in arbitrary space-time dimensions, 
arbitrary signature and
arbitrary signs of the kinetic terms of the field strengths
(see also \cite{ivas}), in the spirit of 
\cite{aeh}. 
We then discuss compactifications of such configurations which lead
to black holes living in an effective Lorentzian space-time (i.e. with
one time coordinate), with particular attention on extremal black holes
with non-vanishing entropy. 
We show that black holes sitting at the two opposite extremes 
of compactification of the physical time have seemingly different
characteristics, namely one is non-singular and has non-vanishing
entropy, and the other is singular and has zero entropy. This situation
is in contrast with what happens for spacelike compactifications, and
leads us to a critical assessment of this enlarged
duality group and the new phases it uncovers.

\section{Generalised Intersection Rules }

In this section we determine how the extremal $p$-branes  of the new phases of
M-theory \cite{huku} intersect to form configurations with zero binding 
energy. The intersection rules will be obtained
in a general framework for arbitrary space-time dimensions, arbitrary signature 
and arbitrary signs of the field strengths along the lines of \cite{aeh}.
The material contained in this section is rather technical, and it is
aimed to satisfy the reader interested in the details of the 
derivation of the brane content of the exotic theories and their 
(generalised) intersection rules. Other readers need not understand
thoroughly these details, and can go directly at the end of the section
where the main results are recollected.

The starting point is the following action\footnote{
We did not take into account in this general approach 
the various Chern-Simons-type terms which occur in the different theories. 
The solutions we will present below can be shown to be 
also  consistent solutions of the full equations 
of motion including such terms.}, where the metric is in the so-called
Einstein frame:
\be
I=\int d^D x \sqrt{|g|} \left\{ R-{1\over 2} (\partial \phi)^2
-{1\over 2} \sum_I {\theta_I \over n_I !} e^{a_I \phi} F_{n_I}^2\right\}.
\label{action}
\ee
The overall sign (and factor) is not relevant. We allow for an 
arbitrary signature which we denote by $(S,T)$, with $S+T=D$, $S$ being 
the number of spacelike dimensions and $T$ the number of 
timelike dimensions \cite{hul2,huku}. 
The $n$-forms' kinetic terms also have an arbitrary 
sign given by $\theta_I=\pm 1$. The equations of motion (EOM)  
and Bianchi identities (BI) are:
\be
R^\mu_\nu={1\over 2}\partial^\mu \phi \partial_\nu \phi +{1\over 2}
\sum_I {\theta_I \over n_I !} e^{a_I \phi} \left[ n_I 
F^{\mu \lambda_2 \dots \lambda_{n_I}}F_{\nu \lambda_2 \dots \lambda_{n_I}}
-{n_I-1\over D-2} \delta^\mu_\nu F_{n_I}^2 \right],
\label{eom1}
\ee
\be
\Box \phi ={1\over 2} \sum_I {\theta_I \over n_I !} a_I e^{a_I \phi}
F_{n_I}^2,
\label{eom2}
\ee
\be
\partial_\mu \left( \sqrt{|g|} e^{a_I \phi} F^{\mu \lambda_2 \dots 
\lambda_{n_I}}\right)=0, \qquad \qquad \partial_{[\mu_1}
F_{\mu_2 \dots \mu_{n_I+1}]}.
\label{eom3}
\ee

Under Hodge duality which interchanges Maxwell EOM  and BI one has:
\be
\begin{array}{lcl}
F_n & \rightarrow & \tilde{F}_{D-n}, \\
n & \rightarrow & \tilde{n}=D-n, \\
a & \rightarrow & \tilde{a}=-a, \\
\theta & \rightarrow & \tilde{\theta}=(-1)^{T+1} \theta.\\
\end{array}
\label{hodge}
\ee
Thus for even $T$ the dual field strength 
has a kinetic term with opposite sign.

Let us now substitute the ans\"atze.
For the metric we take:
\be
ds^2=-\sum_{\alpha=1}^t B_{\alpha}^2 dt_{\alpha}^2 +
\sum_{i=1}^s C_i^2 dy_i^2 +G^2 \eta_{ab} dx^a dx^b,
\label{metric}
\ee
with $\eta_{ab}$ characterizing the overall transverse space-time of 
signature $(S-s, T-t)$ on which all the functions depend, 
and $\delta^a_a=d=S-s+T-t$.

The electric ansatz for a $(s_A, t_A)$-brane is imposed on a
$(s_A+t_A +1)$-form field strength (each brane is labelled by the
index $A$):
\be
F_{\alpha_1\dots \alpha_{t_A}i_1\dots i_{s_A}a}=\epsilon_{\alpha_1\dots 
\alpha_{t_A}}\epsilon_{i_1\dots i_{s_A}} \partial_a E_A.
\label{ean}
\ee
It satisfies trivially the BI.
We actually do not need any magnetic ansatz, since we simply have to
associate a `magnetic' brane to a dual field strength and then translate the
results according to the Hodge duality rules \rref{hodge}.

Let us now rewrite the EOM, enforcing the above ans\"atze \rref{metric}, 
\rref{ean} together
with the `extremality' ansatz \cite{aeh}, namely:
\be
B_1\dots B_t C_1 \dots C_s G^{d-2}=1,
\label{exta}
\ee
which all at once overwhelmingly simplifies the Ricci tensor and 
corresponds to singling out extremal configurations.
The EOM \rref{eom1}-\rref{eom3} then become:
\be
\partial^2 \ln B_\alpha ={1\over 2} \sum_A \gamma_A {\delta^\alpha_A \over
D-2} S_A^2 (\partial E_A)^2,
\label{eo1}
\ee
\be
\partial^2 \ln C_i ={1\over 2} \sum_A \gamma_A {\delta^i_A \over
D-2} S_A^2 (\partial E_A)^2,
\label{eo2}
\ee
\[
\sum_\alpha \partial^a \ln B_\alpha \partial_b \ln B_\alpha +
\sum_i \partial^a \ln C_i \partial_b \ln C_i +
(d-2) \partial^a \ln G \partial_b \ln G + \delta^a_b \partial^2 \ln G =
\]
\be
\qquad =-{1\over 2} \partial^a \phi \partial_b  \phi +{1\over 2}
\sum_A \gamma_A S_A^2 \left[ \partial^a E_A \partial_b  E_A
-{s_A+t_A \over D-2} \delta^a_b (\partial E_A)^2 \right],
\label{eo3}
\ee
\be
\partial^2  \phi = -{1\over 2}\sum_A \gamma_A a_A S_A^2 (\partial E_A)^2,
\label{eo4}
\ee
\be
\partial^a( S_A^2 \partial_a E_A)=0,
\label{eo5}
\ee
where the conventions are:  $v^2=\eta^{ab}v_a v_b$; 
\be
S_A^2=V_A^{-2}e^{a_A \phi}, \qquad \qquad V_A=\prod_{\alpha \parallel A}
B_\alpha \prod_{i\parallel A} C_i.
\ee
The notation $\mu \parallel (\perp) A$ is meant to indicate that
the direction labelled by $\mu$ is longitudinal (perpendicular) to
the brane labelled by $A$.
In the r.h.s. of the  equations \rref{eo1}  we have defined:
\be
\delta^\alpha_A = \left\{\begin{array}{ll}
D-2-s_A-t_A & \mbox{for}\ \alpha \parallel A \\
-(s_A+t_A) & \mbox{for}\ \alpha \perp A
\end{array}\right.
\ee
and similarly for $\delta^i_A$ in eq. \rref{eo2}. Most importantly we have finally:
\be
\gamma_A=\theta_A (-1)^{t_A+1},
\label{gamma}
\ee
instead of having simply $\gamma_A=1$ as in the usual case.

We now enforce the `independence' ansatz \cite{aeh}, namely we take that:
\be
E_A=l_A H_A^{-1}, \qquad \qquad S_A=H_A,
\ee
and that all other functions are products of powers of the $H_A$'s.
This amounts to enforcing the no-force condition between the
intersecting branes.
The Maxwell EOM  \rref{eo5} gives that $\partial^2 H_A=0$, i.e. $H_A$ is
harmonic in $d$-spacetime.
The remaining EOM (together with the first ansatz) give:
\be
\begin{array}{lcl}
\ln B_\alpha&=& - \sum_A \alpha_A {\delta^\alpha_A \over D-2} \ln H_A, \\
\ln C_i &=& - \sum_A \alpha_A {\delta^i_A \over D-2} \ln H_A, \\
\ln G &=& \sum_A \alpha_A {s_A +t_A \over D-2} \ln H_A, \\
\phi &=& \sum_A \alpha_A a_A \ln H_A,
\label{comp}
\end{array}
\ee
where, crucially, we have:
\be
\alpha_A={1\over 2} \gamma_A l_A^2.
\ee
This means that $\alpha_A$ is not necessarily positive from this definition;
rather, it has the same sign of $\gamma_A$, which depends on $t_A$, $\theta_A$
and the electric or magnetic nature of the brane if $T$ is even.

As in \cite{aeh} the last off-diagonal part of the equation \rref{eo3} can be
rewritten as:
\be
\sum_{A,B}\partial^a \ln H_A \partial_b \ln H_B [M_{AB}\alpha_A-\delta_{AB}]
\alpha_B =0,
\ee
with
\be
M_{AB}=\sum_\alpha {\delta^\alpha_A \delta^\alpha_B \over (D-2)^2}+
\sum_i {\delta^i_A \delta^i_B \over (D-2)^2}+(d-2) {(s_A+t_A)(s_B+t_B)
\over (D-2)^2}+{1\over 2} a_A a_B.
\ee
We thus get:
\be
\alpha_A=(M_{AA})^{-1}={D-2\over \Delta_A},
\ee
with 
\be
\Delta_A=(s_A+t_A)(D-2-s_A-t_A)+{1\over 2}a_A^2 (D-2).
\ee
Note that $\Delta_A > 0$ and thus $\alpha_A >0$. If we want to avoid
imaginary $l_A$'s (and thus imaginary field strengths), then all branes
for which $\gamma_A=-1$ are forbidden \cite{huku}.
The powers of the harmonic functions $H_A$'s in the solutions 
are nevertheless exactly the same as for ordinary
$(p,1)$-branes of single-time spacetime.

The last condition, which leads to the intersection rules, is given by $M_{AB}=0$ for
$A\neq B$. It thus leads to:
\be
\bar{s}+\bar{t}={(s_A+t_A)(s_B+t_B)
\over D-2} -{1\over 2} a_A a_B,
\label{wir}
\ee
where $\bar{s}$ and $\bar{t}$ are respectively the number of common spacelike
and timelike directions of the two branes involved\footnote{Note that the
$a_A$'s change sign according to the electric or magnetic nature
of the brane.}.

To summarize the results of this section, we have found that a solution
representing intersecting branes of one of the exotic phases of M-theory
can be simply built taking into account that: a brane with a definite
signature $(s_A,t_A)$ exists only if $\gamma_A\equiv \theta_A (-1)^{t_A+1}=1$;
the intersection rules are given by \rref{wir} and are not restrictive
on $\bar{s}$ and $\bar{t}$ independently; the solution is actually given
by the `harmonic superposition' \cite{tsey} of the single brane solutions,
which involve exactly the same powers of the harmonic function as for the 
branes of the `orthodox' theories.

\section{Black holes and exotic dualities}

In this section we will discuss under which conditions ``exotic" configurations lead upon 
compactification to non-singular extremal black holes. We will also consider the action of the
generalised dualities on configurations of the new phases of M-theory.

We first briefly recall the situation in the  usual phases of M-theory namely type IIA and IIB in 
9+1 dimensions and supergravity in 10+1 dimensions. The corresponding intersection rules are
given by \rref{wir} in the special case  $T=t=\bar{t}=1$ and $t_A=1$ 
for every $A$. From the general solution corresponding to a 
configuration of intersecting branes, 
it is easy to derive the condition under which this configuration after
compactification gives an extremal black hole with non-zero entropy \cite{aeh}. 
The condition to have a non-zero area at $r=0$ is:
\be
{\cal N} =2 {{\bar D -2} \over {\bar D - 3}},
\label{ocon}
\ee
where ${\cal N}$ is the total number of branes in the configuration and ${\bar D}$ is the 
space-time dimensions in which the black hole is living 
(i.e. the overall transverse space directions and the time direction). 
This leads to the well-known result that there are only two cases giving 
regular extremal black holes
namely  ${\cal N}=3$, ${\bar D}=5$ and ${\cal N}=4$, ${\bar D}=4$.  
A key property of these configurations is that the dilaton and the
other moduli approach
a (finite) constant at the horizon $r=0$, and this is precisely the 
reason why the geometry is also regular there, the Bekenstein-Hawking
entropy is finite and moreover it can be matched with
a microscopical derivation (which needs extrapolation from weak
coupling).
All the above ${\bar D}=5$ and ${\bar D}=4$ configurations are related by the
transformations of the U-duality group of, say, type II theories on
$T^5$ and $T^6$ respectively. The mass and the entropy of such black holes can
actually be shown to be U-duality invariant.
The microscopic counting of the entropy of these
black holes \cite{five,four} is nevertheless performed in a particular 
realisation generically involving 
D-branes and momentum.

We now turn to the new phases of M-theory and first state clearly
the setting of the problem. We will be considering only the compactifications
of these `exotic' theories which lead to an effective
(reduced) space-time of Lorentzian signature, i.e. with only one
timelike direction. This means that although the underlying theories
may have unconventional signature, we ask that the effective physics
be conventional, in order to compare with the standard results.

Our focus is thus on configurations of intersecting branes that 
satisfy the following two requirements:
\begin{itemize}
\item 
We restrict ourselves to static configurations, i.e.
the harmonic functions do not depend on any 
time.
All the time directions are generically longitudinal to at least
one brane of the configuration, namely $T=t$.
\item We will compactify on all the timelike directions but one. The non-compact timelike direction
will correspond to the physical time and the expected black hole is living in a spacetime with $S-s$ 
spacelike directions and one timelike direction. 
\end{itemize} 

For a given configuration with ${\cal N}$ branes, one has to 
distinguish between the  branes which are longitudinal to the non-compact 
time  which is chosen as being the physical one and the branes which are 
transverse to this time. 

The branes transverse to the physical time appear to have quite 
unusual features. Indeed, they can be considered from the point of view of the
effective spacetime as some kind of instantons, whose localisation
in the time coordinate is however `smeared' out as far as the present 
solutions are concerned. 
Because of all their longitudinal directions being transverse to the
physical time, their ADM mass turns out to vanish identically.
As a consistency check, one can see that the space-time transverse
to a single such `instantonic' brane reduces indeed to 
flat Minkowski space-time (in the
effective Einstein frame).
Thus, in accordance with the harmonic superposition principle, only the
branes longitudinal to the physical time will contribute to 
the total mass of the configuration.

Focusing now on the entropy, 
if one assume that the configuration considered contains $n$ branes
longitudinal to the physical time and $m$ branes perpendicular to it,
one has obviously the total number of branes which is given by:
\be
{\cal N} = n +  m,
\label{nbrane}
\ee
and one finds that the condition to have a non-zero area is given by:
\be
 n =2 {{\bar D -2} \over {\bar D - 3}},
\label{econ}
\ee
namely the condition depends only on the number $ n$ of branes which ``share" 
the physical
time. The reason for this is that the branes perpendicular to the
`physical' time contribute a factor of 1 to the classical entropy
formula, and thus the total entropy of the compound does not `feel'
them.

The independence with respect to the `instantonic' branes of the 
mass and the entropy has  physical implications.
In some of the new phases of M-theory it is possible to build configurations 
which are singular even if the values of
${\cal N}$ and ${\bar D}$ would have led to regular 
configurations in the ``orthodox" phases. 

To illustrate this fact, we consider an example in the framework of the type ${\rm IIA}_{5+5}$
phase. 
Type ${\rm IIA}_{5+5}$  is a theory with signature (5,5) characterised by the same lagrangian 
as the usual IIA theory, namely all the kinetic terms of the field strengths
have the ``correct"  sign \cite{hul2}.
A configuration of ${\rm IIA}_{5+5}$  with  ${\cal N}=4$ and allowed by the 
intersection rules \rref{wir} is given by (each brane is characterised by $(s_A, t_A)$ the number
of spacelike and timelike directions of its worldvolume):
\be
\begin{array}{rccccccccccc}
{\ } &  t_1 &  t_2 &  t_3 &  t_4 &  t_5 &  | &  y_1 &  y_2 &  x_1 &  x_2 &  x_3 \\
 F1 (1,1) &  X &  - &  - &  - &  - &  | &  X &  - &  - &  - &  - \\
 F1 (1,1) &  - &  X &  - &  - &  - &  | &  - &  X &  - &  - &  - \\
 D4 (0,5) &  X &  X &  X &  X &  X &  | &  - &  - &  - &  - &  - \\
 D4 (2,3) &  - &  - & X & X & X &  | &  X &  X &  - &  - &  - \\
\end{array}
\label{exot}
\ee
For every choice of non-compact physical time this configuration 
corresponds to ${\cal N}=4$ and  ${\bar D}=4$ with $n=2$ and $m=2$. 
It is thus a singular configuration, with
vanishing Bekenstein-Hawking entropy and with the mass being the sum
of two charges, the nature of which depends 
on which time is considered as the physical one (hereafter we will
pick $t_1$).
One can also easily check that the dilaton, as well as all the other
moduli, is singular at $r=0$.

We perform now a series of dualities to map ${\rm IIA}_{5+5}$ onto the 
ordinary IIA theory  in order to
see onto which configuration \rref{exot} is mapped. The series of transformation is
the following (see fig.5 of \cite{hul2}):
\[
\begin{array}{lcccccccl}
 {\rm IIA}_{5+5} & & & & &  & & &  IIA_{9+1} \\
 & & & & & & & & \\
\Downarrow  T_{t_5} & & & & & & & &   \Uparrow T_{t_1} \\
 & & & & & & & & \\
 {\rm IIB}^*_{5+5}& & & & & & &   & {\rm IIB}^*_{9+1}  \\
 & & & & &  & & & \\
\Downarrow S & & & & & & & & \Uparrow S \\
 & & & & & & & & \\
{\rm IIB}^\prime_{5+5} & \Rightarrow & {\rm IIA}_{6+4}& \Rightarrow&
{\rm IIB}_{7+3} & \Rightarrow&{\rm IIA}_{8+2} &\Rightarrow  & 
{\rm IIB}^{\prime}_{9+1} \\ 
& T_{(t_5 \rightarrow  y_3)} & & T_{(t_4 \rightarrow  y_4)} &  
& T_{(t_3 \rightarrow  y_5)} & & T_{(t_2 \rightarrow y_6)} & \\
\end{array}
\]
where $S$ corresponds to a S-duality, $T_{t_i}$ stands 
for a T-duality in the direction $t_i$ which does not affect 
the signature of spacetime, and  $T_{t_i \rightarrow y_j}$ 
corresponds to a T-duality which changes the signature of the spacetime,  
namely mapping a theory compactified on a timelike circle of radius 
$R$ in the $t_i$ direction onto a theory compactified on a 
spacelike circle of radius 
$1/R$, the new spacelike direction being labelled by $y_j$.

Note that the last duality is the trickiest one. While for all the
other ones we can imagine that the direction on which we are
performing the T-duality is compact to begin with and we
stick to the same point in the moduli space, in the case of $t_1$ 
we have to deal with the fact we take it as being the physical time,
and thus we would like to consider it as non-compact both at the 
departure and at the arrival. This entails going, in the variables
pertaining to \rref{exot}, from an infinite radius of compactification
$L$ to a vanishingly small one, $L=0$, which is dual to an infinite radius
after the last T-duality. 

The configuration \rref{exot} becomes after having applied these transformations the following one:

\be
\begin{array}{rccccccccccc}
{\ } &  t_1 &  | &  y_1 &  y_2 &  y_3 & y_4 &  y_5 &  y_6 &  x_1 &  x_2 &  x_3 \\
 NS5  (5,1) &  X &  | &  X  &  - &  X  & X  &  X &  X &  - &  - &  - \\
 D4 (4,1) &  X &  | &  - &  X  &  X  &  X  &  X  &  -  &  - &  - &  - \\
 W  (1,1) &  X &  | &  -  &   -   &  X  &  - &  - &  - &  - &  - &  - \\
 D4 (4,1) &  X  &  |  & X & X &  X &  - &  -  &  X &  - &  - &  - \\
\end{array}
\label{usual}
\ee
This configuration is perfectly regular and leads
upon compactification to the
usual ${\cal N}=4$, ${\bar D}=4$  extremal black hole with non-zero entropy.

We can take as a second example one of the usual configurations
that lead to `countable' black holes, namely those made essentially by
D-branes in type II theories. 
To be more precise, one can consider either a 5-dimensional
black hole made of two kinds of D-branes plus F-strings
(as in the first reference of \cite{five}),
or a 4-dimensional black hole composed uniquely of (four different
kinds of) D-branes, like in \cite{bala}.

In both of the cases above, all the moduli have a constant value at
the horizon $r=0$, thus allowing for a finite value of the entropy.
If we focus on the dilaton, then we can easily see that each $(s,t)$ D-brane
contributes in the following way:
\be
e^{2\phi}=H^{4-s-t \over 2}\times \dots, \label{dcontrib}
\ee
where $H$ is the harmonic function related to the D-brane.
We now perform a timelike T-duality and thus go to type II${}^*$
theories. The (only) timelike direction is common to all the D-branes
in the compound, thus the world-volume of each D-brane will loose one
dimension. This entails that every D-brane will contribute a 
half-power more of its harmonic function to $e^{2\phi}$, thus making now
the latter function explode at the horizon $r=0$. 
A consequence of the explosion of this modulus (and possibly of others)
is that the Bekenstein-Hawking entropy is no longer finite, but rather  
it vanishes, as can be checked independently.

In the particularly striking case of the four-dimensional black hole made
out only of D-branes, one gets after the timelike T-duality a
configuration which has also a vanishing mass, and actually 
the four-dimensional Einstein frame metric is simply the
four-dimensional flat Minkowski one.

It is thus rather easy to find configurations in which the number of
branes longitudinal to the physical time vary when the generalised
dualities are performed. Note however that since in the `orthodox'
theories all the branes have to be longitudinal to the only time,
the entropy is finite whenever the condition \rref{ocon} is met.
To put it in other words, it is possible to go from an `ordinary'
configuration which leads to a black hole
with non-vanishing entropy to an exotic one related to a black hole
with vanishing entropy, but not from an ordinary configuration with
vanishing entropy to an exotic one with non-zero entropy
(always keeping the dimensionality of the black hole fixed).

As a last remark, one could be worried about the possibility 
that, starting with 
a configuration which already counts the right number of
branes longitudinal to the physical time in order to have a 
finite entropy, one could still add more branes perpendicular to the
physical time. 
It turns out however that the generalised 
intersection rules \rref{wir} prevent such a
possibility.

\section{Discussion}
We have analysed in this letter the exotic phases of M-theory
from the point of view of the branes that they contain. 
The first result is that the intersection rules of these branes
\rref{wir} display no surprises with respect to the intersection
rules of the ordinary type II and M-theories \cite{aeh}.
Most notably, the same conclusions can be drawn with respect
to the microscopic interpretation of some of them, namely 
as branes ending on other branes or as branes within other
branes (associated respectively to world-volume gauge theory monopoles and
instantons).

The second result, which might lead to a critical assessment of the new
exotic theories, is that the new dualities, involving T-dualities
along timelike directions, seem to connect smoothly black holes with
different characteristics, namely ones which are non-singular
and have a finite entropy to others which are singular and have a
vanishing entropy.

A heuristic explanation of the mechanism which leads to the 
non-invariance of the entropy is the following. Let us begin by noting that 
quantities such as the mass and the entropy of black holes are usually defined
in a Hamiltonian context, for which a physical time coordinate
has to be inequivocally defined and plays a special r\^ole.
It is thus a legitimate possibility that the procedure of
compactifying the physical timelike direction,
then performing the T-duality, and eventually decompactifying the
same direction can affect these Hamiltonian quantities.
The precise effect of this procedure on the D-branes is that
they become `instantonic' branes, as it can be seen by a simple
string theory argument. This is not straightforward to
see in our classical set-up, since the solutions we 
consider are static, but nevertheless
we argued in the preceding section that these branes give a
vanishing contribution to the ADM mass, a fact which is consistent with
the picture of these branes being events in the transverse 
effective space-time. On these same grounds we do not expect them
to contribute to the entropy of the configuration.

Our precise computation is performed at two points of 
the moduli space of compactification of the physical time,
namely where the original radius and the dual one are respectively
infinite. This is in order to be on safe grounds to perform
a Hamiltonian analysis and derive the black hole thermodynamics.
What happens in the middle, i.e. at finite radius of the physical
time, has to be considered cautiously.

Given that it makes sense to consider a theory in which the
physical time is compact (and thus a moduli space of timelike
compactifications exists),
we are thus left with three options: the entropy does depend on the
radius $L$ of compactification of the physical time, while it is
independent of all the other moduli; the entropy is independent
of $L$, but the entropy and the mass in one theory are mapped to 
other quantities in the dual theory, and vice-versa---thus the theories
might be dual but we do not know how to translate their variables;
the entropy is independent of $L$, and the mismatch on the two
sides means that the transformation is not a duality.

In all cases, the above discussion indicates that timelike
dualities (or, even before, theories with a compact physical time)
lead to rather exotic results. The last option above implies
that the new theories presented in \cite{hul2} are actually
disconnected from the usual phases of M-theory. 
This would mean that the presence of non-unitary fields prevents
these theories to play a r\^ole whatsoever.
The other options
yield a milder assessment, but nevertheless single out
the timelike T-dualities as very particular, seemingly singular, ones.
Note however that acting with T-duality in a timelike direction
which is not taken as the physically relevant one is much closer
to a traditional duality.

Of course we have considered here only solutions with the harmonic
functions depending only on spacelike coordinates (i.e. static solutions
with respect to any time). Moreover, we based our discussion 
on solutions for which we required that the effective space has only
one time. This means that although we consider exotic theories with
possibly several times, we nevertheless stick to a traditional
Lorentzian signature as far as the effective physics is concerned.
A more radical attitude would be to consider on an equal footing
also all the `effective' phases, i.e. all possible compactifications,
but the problems discussed above could actually multiply.

\subsection*{Acknowledgements}

The work of R.~A. is supported by a Golda Meir Fellowship, and also
partially by the Israel Academy of Sciences and Humanities--Centers
of Excellence Programme and the American-Israel Bi-National Science
Foundation.
L.~H. is supported by the grant No. 3980001 from FONDECYT (Chile). 
The Institutional support to the Centro de Estudios Cient\'{\i}ficos de
Santiago
of Fuerza A\'era de Chile and a group of Chilean companies (Empresas CMPC, CGE, 
Codelco, Copec, Minera Collahuasi, Minera Escondida, Novagas, Business
Design Associates, Xerox Chile) is also recognised.\\
R.~A. acknowledges useful criticisms by S.~Elitzur, A.~Giveon and 
E.~Rabinovici, and
L.~H. is grateful to the Theoretical Physics Department of University
Federal of Espirito Santo (Vitoria, Brasil) where part of this work has been
realised.


\begin{thebibliography}{99}


\bibitem{huto} C.~M.~Hull and P.~K.~Townsend, Unity of Superstring 
Dualities, Nucl. Phys. B438 (1995) 109, hep-th/9410167.

\bibitem{witt} E.~Witten, String Theory Dynamics in 
Various Dimensions, Nucl. Phys. B443 (1995) 85,  
hep-th/9503124.

\bibitem{hora} P.~Horava and E.~Witten, Heterotic and Type I String Dynamics
from Eleven Dimensions, Nucl. Phys. B460 (1996) 506, hep-th/9510209.

\bibitem{tasi} J.~Polchinski, TASI Lectures on D-Branes,
hep-th/9611050. 

\bibitem{deuxb} P.~S.~Aspinwall, Some Relationships Between Dualities in String Theory, 
Nucl. Phys. Proc. Suppl. 46 (1996) 30, hep-th/9508154; J.~H.~Schwarz,  
An SL(2,Z) Multiplet of Type IIB Superstrings, 
Phys. Lett. B360 (1995) 13, Erratum-ibid. B364 (1995) 252, hep-th/9508143.


\bibitem{huju} C.~M.~Hull and B.~Julia, Duality and Moduli Spaces for Timelike Reductions,
hep-th/9803239.

\bibitem{crem} E.~Cremmer, I.~V.~Lavrinenko, H.~L\"{u}, C.~N.~Pope, K.~S.~Stelle 
and T.~A.~Tran, Euclidean Signature Supergravities, 
Dualities and Instantons,  hep-th/9803259.

\bibitem{hul1} C.~M.~Hull, Timelike T-Duality, de Sitter Space, Large N Gauge Theories and 
Topological Field Theory, hep-th/9806146.

\bibitem{moor} G.~Moore,  Finite in All Directions, hep-th/9305139; Symmetries and Symmetry breaking
in String Theory, talk given at SUSY 93,  hep-th/9308052.

\bibitem{hul2} C.~M.~Hull, Duality and the Signature of Space-Time, hep-th/9807127.

\bibitem{busc} T.~H.~Buscher, A Symmetry of the String Background Field Equations,
Phys. Lett. B194 (1987) 59; Path Integral Derivation of Quantum Duality in Non Linear Sigma Models, 
Phys. Lett. B201(1988) 466.  

\bibitem{huku} C.~M.~Hull and R.~R.~Khuri, Branes, Times and Dualities, hep-th/9808069.

\bibitem{duff} M.~P.~Blencowe and M.~J.~Duff, Supermembranes and the
Signature of Space-Time, Nucl. Phys. B310 (1988) 387.

\bibitem{tsey} A.~A.~Tseytlin, Harmonic superpositions of M-branes, 
Nucl. Phys. B475 (1996) 149, hep-th/9604035.

\bibitem{kast} J.~P.~Gauntlett, D.~A.~Kastor and J.~Traschen, 
Overlapping Branes in M-Theory, Nucl. Phys.
B478 (1996) 544, hep-th/9604179.

\bibitem{vand} E.~Bergshoeff, M.~de Roo, E.~Eyras, B.~Janssen and 
J.~P.~van der Schaar, Multiple Intersections of D-branes and M-branes,
Nucl. Phys. B494 (1997) 119, hep-th/9612095; Intersections Involving
Monopoles and Waves in Eleven-Dimensions, Class. Quant. Grav. 14 (1997) 2757,
hep-th/9704120.

\bibitem{ivas} V.~D.~Ivashchuk and V.~N.~Melnikov, Sigma-model for
Generalized Composite p-branes, Class. Quant. Grav. 14 (1997) 3001,
hep-th/9705036.

\bibitem{aeh} R.~Argurio, F.~Englert and L.~Houart, Intersection Rules for
p-Branes, Phys. Lett. B398 (1997) 61, hep-th/9701042.

\bibitem{five} A.~Strominger and C.~Vafa, Microscopic Origin of the Bekenstein-Hawking Entropy, 
Phys. Lett. B379 (1996) 99, hep-th/9601029; C.~G.~Callan and J.~M.~Maldacena, 
D-brane Approach to Black Hole Quantum Mechanics, Nucl. Phys. B472 (1996) 591, hep-th/9602043.

\bibitem{four} C.~V.~Johnson, R.~R.~Khuri and R.~C.~Myers, Entropy of 4D Extremal Black Holes,
Phys. Lett. B378 (1996) 78, hep-th/9603061.

\bibitem{bala} V.~Balasubramanian and F.~Larsen, On D-branes and Black
Holes in Four-Dimensions, Nucl. Phys. B478 (1996) 199, hep-th/9604189.

\end{thebibliography}
\end{document}